\documentclass[twocolumn,showpacs,preprintnumbers,amsmath,amssymb,article]{revtex4}

\usepackage{graphicx}
\usepackage{dcolumn}
\usepackage{bm}
\usepackage{mdwlist}
\usepackage{setspace}
\usepackage{natbib}
\usepackage{amssymb}
\usepackage{amsmath}

\begin{document}
\preprint{}

\title{Polarization-dependent ponderomotive gradient force in a standing wave}

\author{P.W. Smorenburg}
\author{J.H.M. Kanters}
\author{A. Lassise}
\author{G.J.H. Brussaard}
\author{L.P.J. Kamp}
\author{O.J. Luiten\footnote{Electronic address: o.j.luiten@tue.nl}}
\affiliation{
Eindhoven University of Technology, Applied Physics, Coherence and Quantum Technology,\\
P.O. Box 513, 5600 MB Eindhoven, The Netherlands
}

\date{\today}

\begin{abstract}
The ponderomotive force is derived for a relativistic charged particle entering an electromagnetic standing wave with a general three-dimensional field distribution and a nonrelativistic intensity, using a perturbation expansion method. It is shown that the well-known ponderomotive gradient force expression does not hold for this situation. The modified expression is still of simple gradient form, but contains additional polarization-dependent terms. These terms arise because the relativistic translational velocity induces a quiver motion in the direction of the magnetic force, which is the direction of large field gradients. Oscillation of the Lorentz factor effectively doubles this magnetic contribution. The derived ponderomotive force generalizes the polarization-dependent electron motion in a standing wave obtained earlier [A.E. Kaplan and A.L. Pokrovsky, Phys. Rev. Lett. $\bm{95}$, 053601 (2005)]. Comparison with simulations in the case of a realistic, non-idealized, three-dimensional field configuration confirms the general validity of the analytical results.
\end{abstract}

\maketitle

\section{Introduction}
The ponderomotive force is a time-averaged force experienced by a charged particle in an oscillating electromagnetic (EM) field that is spatially inhomogeneous. In the standard perturbative approach \cite{Boot,Gaponov}, it is shown that a charged particle in an oscillating EM field attains an oscillatory quiver momentum superimposed on a slowly varying guiding center momentum $\overline{\bm{p}}$. The latter is subject to the classical ponderomotive force $\bm{F}_p$:

\begin{equation}
\frac{d\overline{\bm{p}}}{dt}=\bm{F}_p=-\frac{e^2}{2\epsilon_0 mc\omega^2}\nabla I(\bm{x}),\label{1}
\end{equation}

where $m$ is the mass of the particle, $e$ its charge, $\epsilon_0$ the permittivity, $\omega$ the frequency of the EM field and $I(\bm{x})$ the position-dependent field intensity. The classical ponderomotive force is of gradient form, and always directed toward regions of low field intensity. The ponderomotive force is observed and exploited in a wide range of contexts. In laser-plasma physics, the ponderomotive force drives the formation of laser wakefields that are used for next generation electron accelerators \cite{Tajima,Mangles}. Ion beams are produced by intense laser irradiation of thin foils, in which the ponderomotive force plays an essential role \cite{Denavit,Badziak}. Schemes have been proposed for ponderomotive laser-vacuum acceleration of electrons \cite{Esarey:1,Chaloupka}. In Paul traps, ions are confined by a ponderomotive potential \cite{Paul}. In electron beam diagnostics, the length of electron bunches is measured by sequentially scattering different sections of the bunch using the ponderomotive force of a laser pulse \cite{Siwick,Hebeisen1}.\\

\begin{figure}[h]
\includegraphics[trim = 2.4in 3.3in 0.9in 1.3in, clip, width=0.9\columnwidth]{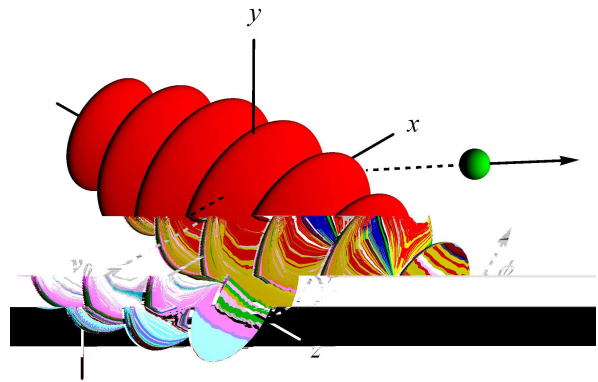}
\caption{\label{fig1}
Charged particle with initial velocity $\bm{v}_0$, which is deflected by the ponderomotive force of a standing EM wave oriented with its nodal planes parallel to the $(x,y)$-plane. The dashed arrow indicates the polarization direction for the case considered in Section \ref{sec.3.1}.}
\end{figure}

The field gradients that can be obtained in a single laser pulse are set by the laser pulse duration longitudinally and the focal spot transversely. For many applications of the ponderomotive force this means that, in order to obtain a sufficiently strong force, field intensities are required that are large enough to cause a relativistic quiver motion (which happens if the normalized amplitude of the vector potential, $a\equiv eA/mc = e\sqrt{2I/(\epsilon_0c)}/(mc\omega)\gtrsim 1$, i.e. $I\gtrsim 2\cdot10^{18}$ W/cm$^2$ for a wavelength of 800 nm). Relativistic field intensities necessitate more complicated descriptions of the average EM force \cite{Bauer,Bituk,Startsev}, or at least restricts the domain of validity of Eq. (\ref{1}) \cite{Dodin,Goreslavskii}. An intermediate situation occurs when an already relativistic particle enters an EM field with nonrelativistic intensity. A relativistic derivation \cite{Kibble} shows that this introduces an additional factor $\sqrt{1+\overline{p}^2/(mc)^2+a^2/2}$ in the denominator of Eq. (\ref{1}), resulting in an accurate description for practical situations \cite{Quesnel}.\\

An alternative to the application of a relativistic laser pulse is the use of a standing wave. In this configuration the nodes and antinodes are spaced on the scale of the wavelength, resulting in large field gradients. For example, a standing wave produced by two counterpropagating EM waves with wavelength $\lambda=800$ nm and a very modest, nonrelativistic peak field intensity of $10^{15}$ W/cm$^2$, already causes ponderomotive forces of the order of $F_p/e\sim1$ GV/m. For this reason, a number of applications of the ponderomotive force have been proposed that take advantage of the large field gradients in a standing EM wave. Hebeisen et al. \cite{Hebeisen2} suggested a table-top standing wave version of the bunch length measurement setup mentioned previously. Following an earlier idea \cite{Fedorov:1}, Balcou proposed a novel X-ray free electron laser based on the wiggling of electrons in the ponderomotive potential of a standing wave \cite{Balcou}. Faure et al. used a standing wave formed by colliding laser pulses to pre-accelerate electrons ponderomotively in a laser-wakefield setup \cite{Faure}, demonstrating that the production of monoenergetic electron beams can be made stable and reproducible in that way \cite{Esarey,Kotaki,Fubiani}. Baum and Zewail proposed to create attosecond electron pulse trains by bunching of an electron beam due to a co-moving ponderomotive beat potential between laser pulses of different frequencies \cite{Baum}.\\

In view of all these important technological applications, a thorough understanding of the ponderomotive force in a standing wave is essential. The scattering of charged particles by a standing EM wave was first described in a quantum-mechanical context by Kapitza and Dirac \cite{Kapitza}, and since then many papers have appeared on this subject \cite{Bucksbaum,Freimund,Fedorov,Batelaan,Li}. Nevertheless, there are only few classical electrodynamical studies on the standing wave ponderomotive force \cite{Bahari,Hartman,Allan,Dodin:1}. Most publications on the ponderomotive force have concentrated on propagating EM waves, establishing the validity of Eq. (\ref{1}) in that context \cite{Boot,Gaponov,Kibble}. Eq. (\ref{1}) is also applied to the standing wave configuration, implicitly assuming that it remains valid in that case as well. In 2005, however, Kaplan and Pokrovsky \cite{Kaplan,Pokrovsky} calculated the time-averaged equation of motion of an electron in a standing wave for a number of field polarizations, and their results showed that the ponderomotive force depends on the polarization. Most notably, the ponderomotive force can even change its direction toward high field regions for certain situations. Clearly, these results are in conflict with the polarization-independent Eq. (\ref{1}) that is commonly used. Kaplan and Pokrovsky did not provide an alternative expression for the ponderomotive force, however.\\

We would now first like to show, on the basis of simple arguments, that it can be understood that the ponderomotive force in a standing wave is polarization-dependent. Consider Fig. \ref{fig1}, showing a particle with charge $q$ and initial velocity $\bm{v}_0$ parallel to the $x$-axis, incident on a standing wave with electric field $\bm{E}$ and magnetic field $\bm{B}$. The wave is oriented with its nodal planes parallel to the $(x,y)$-plane, so that the spatial variation of the field is much more rapid in the $z$-direction than in the transverse direction. This is the typical system considered in this paper. When the particle enters the EM field, it will start to quiver in the polarization direction in response to the oscillating electric force $q\bm{E}$. This 'electric quiver', combined with the Lorentz force equation, leads to the well-known average force, Eq. (\ref{1}), independent of the polarization direction. However, the incident particle will also quiver in response to the magnetic force $\approx q(\bm{v}_0\times\bm{B})$. This 'magnetic quiver' can be comparable in magnitude to the electric quiver for relativistic particles. Because the magnetic quiver is in the $z$-direction, the particle samples a large field gradient, leading to an additional contribution to the ponderomotive force that is comparable to the electric one. And since the magnetic force and hence the amplitude of the magnetic quiver depend on the angle between $\bm{v}_0$ and $\bm{B}$, the magnetic contribution is dependent on the polarization direction.\\

In this paper, the ponderomotive force is derived for a relativistic particle entering a nonrelativistic standing wave with a general three-dimensional field distribution. It is shown that indeed Eq. (\ref{1}), or the relativistic equivalent, does not hold for this situation. This may have important implications for experiments and proposals based on the standing wave ponderomotive force. The main result of this paper, Eq. (\ref{22}), shows that the modified ponderomotive force is still of simple gradient form, but contains additional polarization-dependent terms. We identify the polarization-dependent contribution due to the magnetic quiver. An additional relativistic effect effectively doubles this magnetic contribution. We thus generalize the results of Kaplan and Pokrovsky, which follow naturally from our ponderomotive force expression. This paper is structured as follows. In Section \ref{sec.2}, the polarization-dependent ponderomotive gradient force in a standing wave, Eq (\ref{22}), is derived using a perturbation expansion method. The origin of the additional polarization-dependent terms is discussed. Next, Eq. (\ref{22}) is applied in Section \ref{sec.3} to calculate the averaged equation of motion of an electron in specific standing wave geometries, reproducing the results obtained by Kaplan and Pokrovsky. In Section \ref{sec.4}, we validate our ponderomotive force expression by testing it against numerical simulations of electron trajectories in a realistic, non-idealized field configuration.

\section{The polarization-dependent ponderomotive force \label{sec.2}}
\subsection{Assumptions}
First consider the idealized case of two plane EM waves of equal frequency $\omega$ that counterpropagate along the $z$-axis and add to form a standing wave. In the Coulomb gauge, the vector potential $\bm{A}$ of this ideal standing wave is then purely harmonic in time $t$ and position $z$, i.e. $\left(\partial^2/\partial t^2+\omega^2\right)\bm{A}=\bm{0}$ and $\left(\partial^2/\partial z^2+k^2\right)\bm{A}=\bm{0}$, where $k=\omega/c$ with $c$ the speed of light. Furthermore, the vector potential satisfies $A_z=0$ and $\nabla_\perp A_i=\bm{0}$, where $i=x,y$.\\

In practical applications, however, an EM standing wave differs from this idealized situation in two ways. First, the standing wave has a finite transverse extent, leading to a small but nonzero transverse gradient $\nabla_\perp A_i$ and a small longitudinal component $A_z$. These two quantities are related by the gauge condition $\nabla\cdot\bm{A}=0$, and scale analysis of the latter shows that, symbolically,
\begin{equation}
\frac{A_z}{A_\perp}\sim\frac{\nabla_\perp}{\partial/\partial z}\sim\epsilon.\label{5}
\end{equation}
Here, $\epsilon\ll1$ is a small parameter measuring the magnitude of the field inhomogeneity, and will be used as the expansion parameter in the derivation that follows. For example, in Gaussian laser beams focused to a waist of size $w_0$, this parameter is $\epsilon\sim(kw_0)^{-1}$.

Secondly, the standing wave has a finite lifetime, so that the vector potential is only quasi-monochromatic:
\begin{equation}
\frac{\partial^2\bm{A}}{\partial (\omega t)^2}+\bm{A}=O(\delta A);\hspace{0.3cm}\frac{\partial^2\bm{A}}{\partial (kz)^2}+\bm{A}=O(\delta A),\label{4}
\end{equation}
where $\delta\ll 1$ is another small parameter measuring the monochromaticity. For a standing wave produced by counterpropagating laser pulses of temporal length $\sigma$, for example, this parameter is $\delta\sim(\omega\sigma)^{-1}$.\\

In addition, in this paper the EM field is assumed to be of nonrelativistic intensity, which means that
\begin{equation}
\frac{eA}{mc}\equiv a\ll 1. \label{3}
\end{equation}
For reasons of clarity, for the moment it is assumed that $a\sim\epsilon$. However, the derivation below can be generalized straightforwardly to other field strengths such as $a\sim\epsilon^2$ or $a\sim\epsilon^{1/2}$, leading to the same result. Appendix \ref{app.2} gives a short description of the generalized derivation.

\subsection{Perturbation expansions \label{sec2.2}}

Before considering the dynamics of a charged particle in a standing wave in detail, let us first determine what time scales are involved. First, there is the time scale of the quiver motion, which is the optical time scale $\omega^{-1}$. Second, there is the time scale on which the motion of the guiding center changes. Substitution of an ideal standing wave $\bm{A}=A_0\bm{e}_x\cos kz\sin\omega t$ in Eq. (\ref{1}) and integrating yields oscillatory motion in the $z$-direction, with a typical frequency $\Omega=eA_0\omega/(\sqrt{2}mc)$. Thus the guiding center motion in the $z$-direction changes on a second, longer time scale $(a\omega)^{-1}\sim(\epsilon\omega)^{-1}$. Finally, in a realistic standing wave, the nonzero transverse field gradient causes transverse ponderomotive forces, which in view of Eqs. (\ref{1}) and (\ref{5}) are weaker than the longitudinal ponderomotive forces by a factor $\epsilon$. Therefore the transverse guiding center motion changes on a third, still longer time scale $(\epsilon^2\omega)^{-1}$.
\\

Having established the three time scales of the problem, next consider the equations of motion of a charged particle in the standing wave \cite{Griffiths}:
\begin{eqnarray}
\frac{d}{d(\omega t)}\left(\frac{\bm{p}}{mc}+\frac{e\bm{A}}{mc}\right)&=&\frac{1}{\gamma}\left(\frac{\lambda\nabla}{2\pi}\frac{e\bm{A}}{mc}\right)\cdot\frac{\bm{p}}{mc};\label{6}\\
\frac{d(k\bm{x})}{d(\omega t)}&=&\frac{1}{\gamma}\frac{\bm{p}}{mc},\label{7}
\end{eqnarray}
in which $\gamma=\sqrt{1+p^2/(mc)^2}$ is the Lorentz factor, and the dyadic notation $\nabla\bm{A}$ has been used \cite{Morse}. Eqs. (\ref{6})-(\ref{7}) have been made dimensionless by dividing the usual equations by $mc\omega$ and $c$ respectively. Below, these equations are solved by expressing the various quantities in perturbation expansions in terms of $\epsilon$. Subsequently, terms of like order in $\epsilon$ will be collected and equated \cite{Nayfeh}. We use the symbol '$Ord$' to denote 'on the order of', i.e. $f=Ord(\epsilon^i)$ means $0<\lim_{\epsilon\downarrow 0}f/\epsilon^i<\infty$, in distinction with $f=O(\epsilon^i)$ which is equivalent to $0\leq\lim_{\epsilon\downarrow 0}f/\epsilon^i<\infty$. Superscripts in parentheses denote the order of the terms.\\

First, the momentum is expanded as
\begin{eqnarray}
\bm{p}&=&\sum_{i=0}^{\infty} \bm{p}^{(i)};\label{8}\\
\bm{p}^{(i)}&=&Ord(\epsilon^i).\nonumber
\end{eqnarray}
Note that $\bm{p}^{(0)}$ would be the momentum in the absence of the standing wave field, that is, the initial momentum, since $\bm{p}\rightarrow\bm{p}^{(0)}$ as $a\sim\epsilon\downarrow 0$. Next, each order $\bm{p}^{(i)}$ is decomposed into a slowly varying guiding center part $\overline{\bm{p}}^{(i)}=\langle\bm{p}^{(i)}\rangle$ and a rapidly varying quiver part $\widetilde{\bm{p}}^{(i)}=\bm{p}^{(i)}-\langle\bm{p}^{(i)}\rangle$. Here, $\langle\cdot\rangle$ denotes time-averaging on the time scale $\omega^{-1}$. Upon substitution of this decomposition in the left-hand side of Eq. (\ref{6}), each term is differentiated with respect to $\omega t$. This preserves the order of magnitude of the fast quantities $\widetilde{\bm{p}}^{(i)}$, since these vary on the time scale $\omega^{-1}$, i.e. $d\widetilde{\bm{p}}^{(i)}/d(\omega t)=Ord(\epsilon^i)$. However, from the discussion above, the slow quantities $\overline{p}_z^{(i)}$ and $\overline{\bm{p}}_\perp^{(i)}$ vary over longer time scales, so that differentiation increases their order according to $d\overline{p}_z^{(i)}/d(\omega t)=Ord(\epsilon^{i+1})$ and $d\overline{\bm{p}}_\perp^{(i)}/d(\omega t)=Ord(\epsilon^{i+2})$ respectively. More formally, these order relations may be established using the multiple scale technique \cite{Nayfeh}, considering $p_z$ a function of the two variables $\omega t$ and $\epsilon\omega t$ and considering $\bm{p}_\perp$ a function of the two variables $\omega t$ and $\epsilon^2\omega t$. Multiple scale analysis has been applied in a relativistic derivation of the ponderomotive force in propagating EM radiation \cite{Startsev}.\\

Next, the expansion (\ref{8}) is substituted in the reciprocal Lorentz factor $1/\gamma$. Extracting the zeroth order part $\gamma^{(0)}=\sqrt{1+\left(p^{(0)}\right)^2/(mc)^2}$, this gives
\begin{eqnarray}
\frac{1}{\gamma}&=&\frac{1}{\gamma^{(0)}\sqrt{1+\left(mc\gamma^{(0)}\right)^{-2}\left(2\bm{p}^{(0)}\cdot\bm{p}^{(1)}+O(\epsilon^2)\right)}}=\nonumber\\
&=&\frac{1}{\gamma^{(0)}}-\frac{\bm{p}^{(0)}\cdot\bm{p}^{(1)}}{(mc)^2\left(\gamma^{(0)}\right)^3}+O(\epsilon^2)\equiv\nonumber\\
&\equiv&\left(\frac{1}{\gamma}\right)^{(0)}\!\!\!\!\!\!+\left(\frac{1}{\gamma}\right)^{(1)}\!\!\!\!\!\!+O(\epsilon^2).\label{10}
\end{eqnarray}
The quantity $(1/\gamma)^{(1)}$ is the first-order time-dependent variation of the reciprocal Lorentz factor with respect to the constant value $(1/\gamma)^{(0)}$. As will be shown below, the associated variation in the relativistic mass $\gamma m$ of the particle leads to an additional contribution in the final ponderomotive force expression.\\

Finally, below it will be required to take the time average of expressions involving powers of $\bm{A}$ or its derivatives. These time averages need to be taken along the trajectory of the particle; that is, in the average $\langle\bm{A}\left(\bm{x}(t),t\right)\rangle$, the vector potential is to be evaluated at $\bm{x}=\bm{x}(t)$. To bring out this position dependence explicitly, the position is also expanded in a perturbation expansion in terms of $\epsilon$:
\begin{eqnarray}
\bm{x}&=&\sum_{i=0}^{\infty} \bm{x}^{(i)};\label{11}\\
\frac{d\bm{x}^{(i)}}{dt}&=&Ord\left(\bm{p}^{(i)}\right)=Ord(\epsilon^i).\nonumber
\end{eqnarray}
Again, each order $\bm{x}^{(i)}$ is decomposed into a slowly varying guiding center part $\overline{\bm{x}}^{(i)}=\langle\bm{x}^{(i)}\rangle$ and a rapidly varying quiver part $\widetilde{\bm{x}}^{(i)}=\bm{x}^{(i)}-\langle\bm{x}^{(i)}\rangle$. Then the vector potential can be expanded in a Taylor series around $\bm{x}=\sum\overline{\bm{x}}^{(i)}\equiv\overline{\bm{x}}$,
\begin{equation}
\bm{A}\left(\bm{x}(t),t\right)=\overline{\bm{A}}+\widetilde{z}^{(1)}(t)\frac{\partial\overline{\bm{A}}}{\partial z}+O(\epsilon^3).\label{13}
\end{equation}
Here and in the remainder of the paper, an overbar on the vector potential denotes evaluation at the guiding center position, i.e. $\overline{\bm{A}}\equiv\bm{A}\left(\overline{\bm{x}},t\right)$. In writing the series in Eq. (\ref{13}), it has been anticipated that $\widetilde{\bm{x}}^{(0)}=\bm{0}$, as will be shown below.

\subsection{Order-by-order solution of equation of motion}
Substitution of the expansions (\ref{8})-(\ref{13}) in the equations of motion (\ref{6})-(\ref{7}), and collecting terms of equal order in $\epsilon$, results in two equations at each order of $\epsilon$. These order equations are listed in Appendix \ref{app.1}. Order-by-order solution, balancing in each equation the averaged parts and the oscillating parts separately, yields the zeroth order quantities
\begin{eqnarray}
\overline{\bm{p}}^{(0)}&=&\bm{p}_0;\label{14}\\
\widetilde{\bm{p}}^{(0)}&=&\bm{0};\label{15}\\
\frac{d\overline{\bm{x}}^{(0)}}{dt}&=&\frac{\bm{p}_0}{m\gamma^{(0)}};\label{16}\\
\widetilde{\bm{x}}^{(0)}&=&\bm{0},\label{17}
\end{eqnarray}
in which $\bm{p}_0$ is the initial momentum. As expected, at zeroth order (that is, in the limit $a\sim\epsilon\downarrow 0$ where both the field strength and field inhomogeneity are zero) the motion is equal to what it would be if the EM field were absent. For the first order quantities, it is found that
\begin{eqnarray}
\widetilde{\bm{p}}_\perp^{(1)}&=&-e\bm{A}_\perp;\label{18}\\
\frac{d\widetilde{p}_z^{(1)}}{dt}&=&\frac{e}{m\gamma^{(0)}}\frac{\partial\overline{\bm{A}}_\perp}{\partial z}\cdot\bm{p}_{0\perp};\label{19}\\
\frac{d\widetilde{z}^{(1)}}{dt}&=&\frac{1}{m\gamma^{(0)}}\!\!\left(\widetilde{p}_z^{(1)}\!+\!p_{0z}\frac{e\bm{A}_\perp\!\cdot\!\bm{p}_{0\perp}-p_{0z}\widetilde{p}_z^{(1)}}{\left(mc\gamma^{(0)}\right)^2}\right)\!;\label{20}\\
\frac{d\overline{z}^{(1)}}{dt}&=&\frac{\overline{p}_z^{(1)}}{m\gamma^{(0)}}.\label{20a}
\end{eqnarray}
Eq. (\ref{18}) expresses the well-known result that, in an oscillating EM field, at lowest order the quiver momentum balances the vector potential, such that the canonical momentum $\bm{p}+e\bm{A}$ is conserved. Eq. (\ref{20a}) will be used in Section \ref{sec.3} for the description of the guiding center motion. Eqs. (\ref{19})-(\ref{20}) describe the quiver motion in the direction normal to the plane of polarization of the standing wave, which is the direction of strong field gradient. This is the magnetic quiver motion described in the Introduction. Time differentiation of Eq. (\ref{20}) and substitution of Eq. (\ref{19}) yields
\begin{eqnarray}
\frac{d^2\widetilde{z}^{(1)}}{dt^2}=\frac{e}{(m\gamma^{(0)})^2}\left((1-\beta_{0z}^2)\frac{\partial\overline{\bm{A}}_\perp}{\partial z}\!+\!\frac{\beta_{0z}}{c}\frac{d\bm{A}_\perp}{dt}\right)\!\cdot\bm{p}_{0\perp},\nonumber\\
\label{20b}
\end{eqnarray}
where $\bm{\beta}_0=\bm{p}_0/(mc\gamma^{(0)})$ is the initial velocity divided by $c$. We now restrict to the situation that $\beta_{0z}$ is sufficiently small so that the second term in parentheses in Eq. (\ref{20b}) is negligible, which is the case if $\beta_{0z}\ll1$. Then, in addition, using Eq. (\ref{4}) the full time derivative may be written $d^2/dt^2=\left(\partial/\partial t+\bm{v}\cdot\nabla\right)^2=-\omega^2+O(\epsilon,\delta,\beta_{0z})$, so that double time integration yields
\begin{equation}
\widetilde{z}^{(1)}=-\frac{e}{\left(m\gamma^{(0)}\right)^2\omega^2}\frac{\partial\overline{\bm{A}}_\perp}{\partial z}\cdot\bm{p}_{0\perp}\left[1+O(\delta,\beta_{0z})\right].\label{21}
\end{equation}
The bracketed factor expresses the error introduced by the integration. Eq. (\ref{21}) clearly shows that the amplitude of the magnetic quiver motion is polarization-dependent: if $\bm{p}_{0\perp}$ is parallel to $\bm{A}_\perp$, this amplitude may be substantial, while for $\bm{p}_{0\perp}$ perpendicular to $\bm{A}_\perp$, it vanishes at first order. This reinstates the argument made in the Introduction: if $\bm{p}_{0\perp}\parallel\bm{A}_\perp$, the momentum is largely perpendicular to the magnetic field, resulting in a substantial magnetic force and quiver amplitude. Conversely, if $\bm{p}_{0\perp}\perp\bm{A}_\perp$, the momentum is largely parallel to the magnetic field, with vanishing magnetic force and quiver amplitude.\\

As a final step, we substitute Eqs. (\ref{14})-(\ref{18}) and (\ref{21}) in the right-hand sides of the remaining order equations (\ref{A7})-(\ref{A8}), and take the time average of these equations. Then the left-hand sides reduce to the rate of change of the first order guiding center momentum, $d\overline{\bm{p}}^{(1)}/dt$. The right-hand sides reduce to a single gradient:
\begin{align}
\frac{d\overline{\bm{p}}^{(1)}}{dt}&\approx-\frac{e^2}{2m\gamma^{(0)}}\cdot\label{22}\\
\cdot\nabla&\left\langle\overline{\bm{A}}_\perp^2-\left(\bm{\beta}_{0\perp}\cdot\overline{\bm{A}}_\perp\right)^2+\left(\frac{\partial\left(\bm{\beta}_{0\perp}\cdot\overline{\bm{A}}_\perp\right)}{\partial(kz)}\right)^2\right\rangle.\nonumber
\end{align}
This is the polarization-dependent ponderomotive force in a nonrelativistic standing wave for a particle with $\beta_{0z}\ll1$; it is the main result of this paper. The approximate sign expresses a relative error of the order of $\beta_{0z}+\delta$. In the limit $\bm{\beta}_{0\perp}\rightarrow 0$, Eq. (\ref{22}) reduces to the well-known polarization-independent ponderomotive force, Eq. (\ref{1}), with the relativistic factor $\gamma^{(0)}\approx\sqrt{1+\overline{p}^2/(mc)^2+a^2/2}$ included in the denominator. For $\bm{\beta}_{0\perp}\neq0$, however, the two polarization-dependent terms of Eq. (\ref{22}) become significant.\\

The term in Eq. (\ref{22}) proportional to $\left(\partial\left(\bm{\beta}_{0\perp}\cdot\overline{\bm{A}}_\perp\right)/\partial kz\right)^2$ originates from including the magnetic quiver motion, Eq. (\ref{21}), in the Taylor expansion of the vector potential, Eq. (\ref{13}). It accounts for the fact that the $z$-position oscillates in phase with the temporal oscillation of the field. Therefore, when a field gradient in the $z$-direction is present, the particle systematically samples higher fields at selected phases of the electric quiver motion in the direction of $\bm{A}_\perp$, and lower fields at other phases. The induced nonzero average force is negligible in most applications of the ponderomotive force. But in a standing wave the derivative in $\partial^2/\partial (kz)^2$ is of order unity, so that this term is comparable to the other terms in Eq. (\ref{22}).

The origin of the term in Eq. (\ref{22}) proportional to $\left(\bm{\beta}_{0\perp}\cdot\overline{\bm{A}}_\perp\right)^2$ can be traced back to including the first order term $\left(1/\gamma\right)^{(1)}$ in the expansion of the Lorentz factor, Eq. (\ref{10}), rather than approximating $\gamma\approx\gamma^{(0)}$ throughout. This takes into account that the relativistic mass $\gamma m$ of the particle oscillates around the value $\gamma^{(0)}m$ in phase with the oscillation of the field. Therefore, during its primary quiver motion in the direction of $\bm{A}_\perp$, the particle is systematically heavier at selected phases of the quiver motion, and lighter at other phases, leading to a nonzero effective force when a field gradient is present.

\section{Wiggling motion in a standing wave \label{sec.3}}
The last term of the $z$-component of Eq. (\ref{22}) may be rewritten by performing the $z$-differentiations and using Eq. (\ref{4}), after which Eq. (\ref{22}) becomes
\begin{equation}
\frac{d\overline{p}_z^{(1)}}{dt}\approx-\frac{e^2}{2m\gamma^{(0)}}\frac{\partial}{\partial z}\left\langle\overline{\bm{A}}_\perp^2-2\left(\bm{\beta}_{0\perp}\cdot\overline{\bm{A}}_\perp\right)^2\right\rangle\label{23}.
\end{equation}\\
We now evaluate Eq. (\ref{23}) for the linearly and circularly polarized standing waves considered by Pokrovsky and Kaplan.

\subsection{Linear polarization \label{sec.3.1}}
Let the standing wave be produced by two counterpropagating plane waves of equal amplitude and frequency that are collinearly polarized in the direction $\bm{e}_p=\bm{e}_x\cos\phi+\bm{e}_y\sin\phi$, as is indicated in Fig. \ref{fig1} by the dashed arrow. Then the vector potential is $\bm{A}=A_0\bm{e}_p\cos kz\sin\omega t$. Suppose that a charged particle enters the standing wave parallel to the $x$-axis with initial velocity $\beta_0$. Differentiating Eq. (\ref{20a}), and substituting Eq. (\ref{23}), gives the equation of motion for the guiding center of the particle in the $z$-direction:
\begin{equation}
\frac{d^2k\overline{z}}{d(\omega t)^2}-a_0^2\frac{1-2\beta_0^2\cos^2\phi}{4\left(\gamma^{(0)}\right)^2}\sin(2k\overline{z})\approx 0,\label{24}
\end{equation}
in which $a_0=eA_0/(mc)$. Eq. (\ref{24}) shows that the guiding center makes pendulum-like oscillations in the $z$-direction (it wiggles in the ponderomotive potential), with equilibrium points at $k\overline{z}=n\pi/2$ and a small-amplitude frequency $\Omega=a_0\omega\sqrt{\left|1-2\beta_0^2\cos^2\phi\right|}/(\sqrt{2}\gamma^{(0)})$. \\

First, consider polarization parallel to the initial velocity, i.e. $\cos\phi=1$. Then Eq. (\ref{24}) reduces to Eq. (40) of Ref. \cite{Pokrovsky} after rewriting $\beta_0=p_0/(mc\gamma^{(0)})$. As was noted in Ref. \cite{Pokrovsky}, the most striking feature of this configuration is that the guiding center oscillations vanish for $\beta_0=1/\sqrt{2}$; in terms of the guiding center motion, the standing wave is invisible to the particle for this value of initial velocity. When $\beta_0$ is increased above $1/\sqrt{2}$, the stable and unstable equilibrium points of Eq. (\ref{24}) reverse their positions, i.e. the ponderomotive force changes direction towards high field regions. This is the relativistic reversal described in Ref. \cite{Pokrovsky}.

For polarization perpendicular to the initial velocity of the particle, $\cos\phi=0$. Then Eq. (\ref{24}) becomes identical to Eq. (52) of Ref. \cite{Pokrovsky}. For this polarization, the magnitude of the ponderomotive force and the wiggling frequency are independent of the initial velocity, and the relativistic reversal effect is absent.

\subsection{Circular polarization \label{sec.3.2}}
If the standing wave is produced by two counterpropagating plane waves that are circularly polarized with opposite helicities, the vector potential is equal to $\bm{A}=A_0\cos kz\left(\bm{e}_x\cos\omega t+\bm{e}_y\sin\omega t\right)$. Thus the standing wave has equally spaced nodes and antinodes along the $z$-axis, while locally the field direction rotates around the $z$-axis with time. Again combining Eqs. (\ref{20a}) and (\ref{23}) and substituting the vector potential now gives
\begin{equation}
\frac{d^2k\overline{z}}{d(\omega t)^2}-\frac{a_0^2}{2\left(\gamma^{(0)}\right)^4}\sin(2k\overline{z})\approx 0,\label{26}
\end{equation}
which equals Eq. (56) of Ref. \cite{Pokrovsky} taking into account that the field amplitude used there is $A_0/\sqrt{2}$. Thus, in this configuration the equilibrium points are the same as in the linearly polarized case described by Eq. (\ref{24}) for $\cos\phi=0$, although the magnitude of the ponderomotive force is a factor $\left(\gamma^{(0)}\right)^2$ weaker.\\

For counterpropagating circularly polarized waves of equal helicity, the vector potential reads $A_0\sin\omega t\left(\bm{e}_x\sin kz+\bm{e}_y\cos kz\right)$. In this case, the standing wave is polarized in a helix along the $z$-axis. Eqs. (\ref{20a}) and (\ref{23}) yield
\begin{equation}
\frac{d^2k\overline{z}}{d(\omega t)^2}+\frac{a_0^2\beta_0^2}{2\left(\gamma^{(0)}\right)^2}\sin(2k\overline{z})\approx 0,\label{27}
\end{equation}
which is the same as Eq. (62) of Ref. \cite{Pokrovsky} after rewriting $\beta_{0}=p_{0}/(mc\gamma^{(0)})$ and taking into account the different definition of the field amplitude. As was noted in Ref. \cite{Pokrovsky}, the field intensity is homogeneous along the $z$-axis, so that Eq. (\ref{1}) predicts zero ponderomotive force. However, the modified ponderomotive force expression, Eq. (\ref{22}), shows that this force is nonzero, so that Eq. (\ref{27}) still yields wiggling motion of the guiding center.\\

In summary, all the equations of motion (\ref{24})-(\ref{27}) can be derived from a single ponderomotive force expression, Eq. (\ref{22}). We have therefore generalized the results of Kaplan and Pokrovsky, who started from the Lorentz force equation for each individual case. Moreover, Eq. (\ref{22}) also gives the transverse component of ponderomotive force, which was not considered in Ref. \cite{Pokrovsky}.\\

\section{Comparison with simulations \label{sec.4}}
In this section, Eq. (\ref{22}) is tested against numerical simulations of electron trajectories in a realistic, non-idealized standing wave field. We have used the GPT code which uses an embedded fifth order Runge-Kutta method with adaptive stepsize control \cite{gpt,url}. For comparison with the simulations, Eq. (\ref{22}) is needed in terms of the electric field rather than the vector potential. From Eq. (\ref{4}), the potential is approximately harmonic in time, so that $\langle\overline{\bm{A}}^2_\perp\rangle\approx\omega^{-2}\langle\overline{\bm{E}}^2_\perp\rangle$. Therefore $\overline{\bm{A}}_\perp$ may be effectively replaced by $\overline{\bm{E}}_\perp/\omega$ in Eq. (\ref{22}).\\

We consider again the configuration shown in Fig. \ref{fig1}, this time with two identical Gaussian laser beams in the fundamental mode that counterpropagate along the $z$-axis, have a central wavelength $\lambda=800$ nm and a peak intensity $I_0=2.0\cdot10^{14}$ W/cm$^2$, and are focused in a common waist of size $w_0=12.5$ $\mu$m at $z=0$. For these parameters, $\epsilon\sim(kw_0)^{-1}=0.01$ and $a=0.01$ so that the theory of Section \ref{sec.2} is valid. The beams are assumed to be pulsed with a Gaussian pulse shape of length $\sigma=85$ fs, and timed such that the centers of the pulses coincide at $z=0$ at time $t=0$. This configuration yields a standing wave around the origin with length $\sim c\sigma$ and width $\sim w_0$, and with a lifetime $\sim\sigma$ around $t=0$. Furthermore, we assume that the pulses are collinearly polarized in the direction $\bm{e}_p=\bm{e}_x\cos\phi+\bm{e}_y\sin\phi$, similar to the case considered in Section \ref{sec.3.1}. In each of the following cases, the trajectory is calculated of an electron entering with initial velocity $\bm{\beta}_0=\beta_0\bm{e}_x$ for $t\ll-\sigma$, and with the initial position chosen such that, at $t=0$, the position of the electron would be $(x,y,z)=(0,w_0/2,\lambda/8)$ in absence of the laser fields. Then the electron meets the standing wave close to the origin, has interaction with it for some time $\sim\sigma$, and leaves the interaction region in a deflected direction. The initial position has been chosen such that the electron samples the highest available field gradients both in the $z$-direction and in the perpendicular direction, in order to maximize the ponderomotive effects.\\

\begin{figure}[p]
\includegraphics[trim = 0.1in 0in 0.6in 0in, clip, width=0.9\columnwidth]{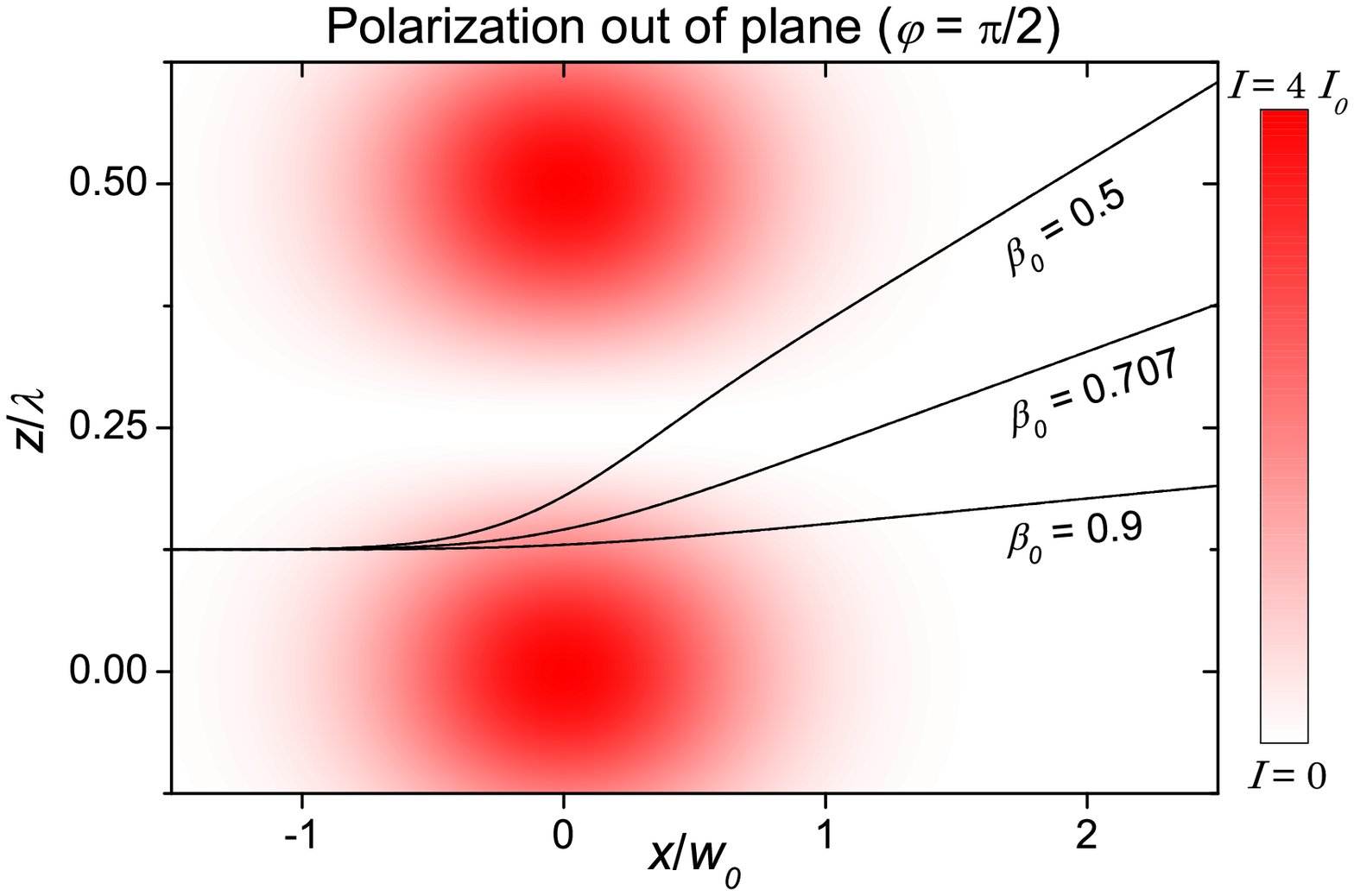}
\caption{\label{fig2}
Trajectory of an electron incident on a standing wave that is polarized in the $y$-direction (out of plane), for three different initial velocities $\beta_0$. The color map shows the field intensity of the standing wave in the $(z,x)$-plane at time $t=0$. For each initial velocity, the plot actually includes three trajectories, calculated with the methods GPT, OLD, and NEW respectively; in each case all three lines overlap to within the linewidth.}
\end{figure}

\begin{figure}[p]
\begin{tabular}{c}
\includegraphics[trim = 0.1in 0in 0.6in 0in, clip, width=0.9\columnwidth]{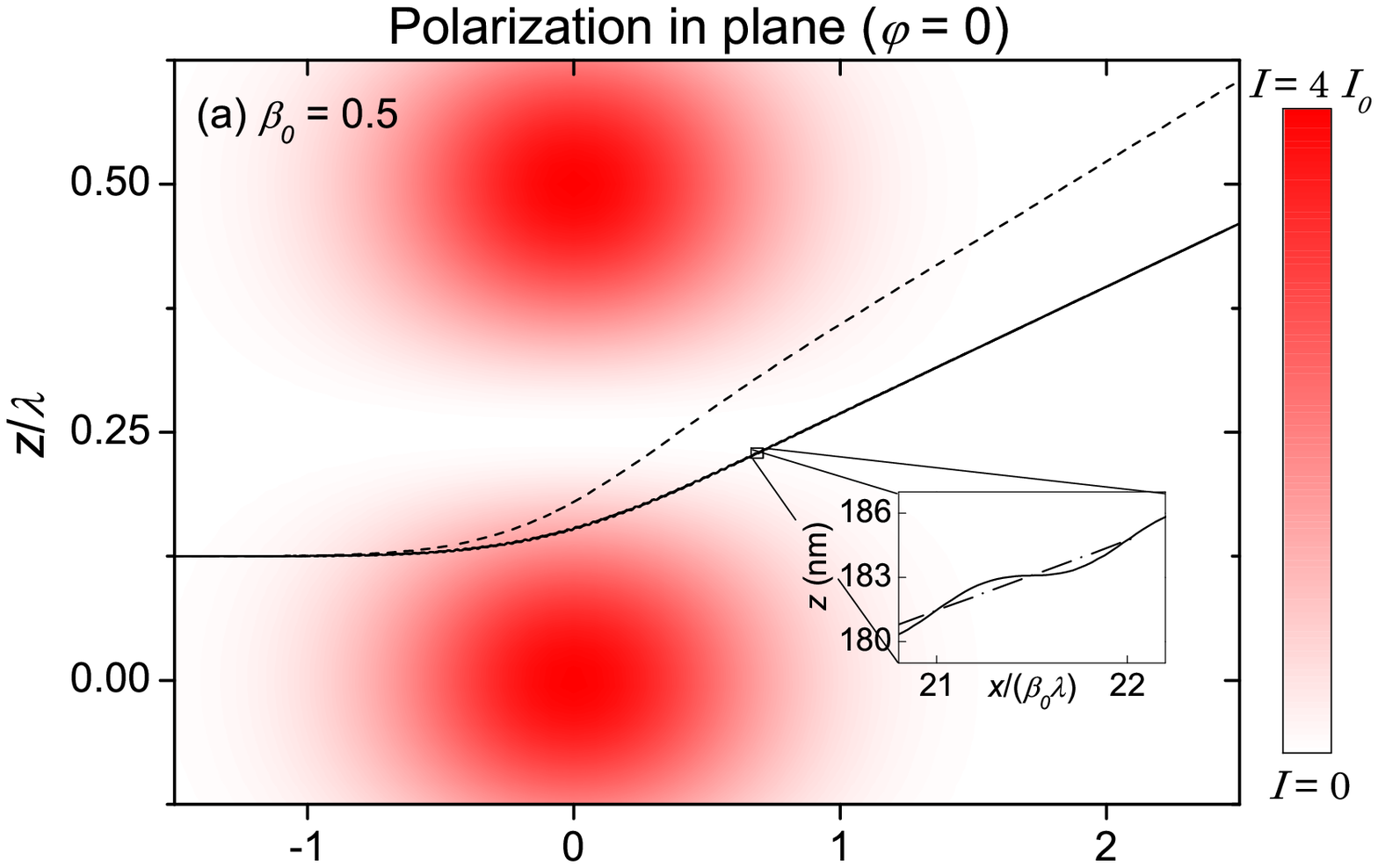}\\[-2mm]
\includegraphics[trim = 0.1in 0in 0.6in 0in, clip, width=0.9\columnwidth]{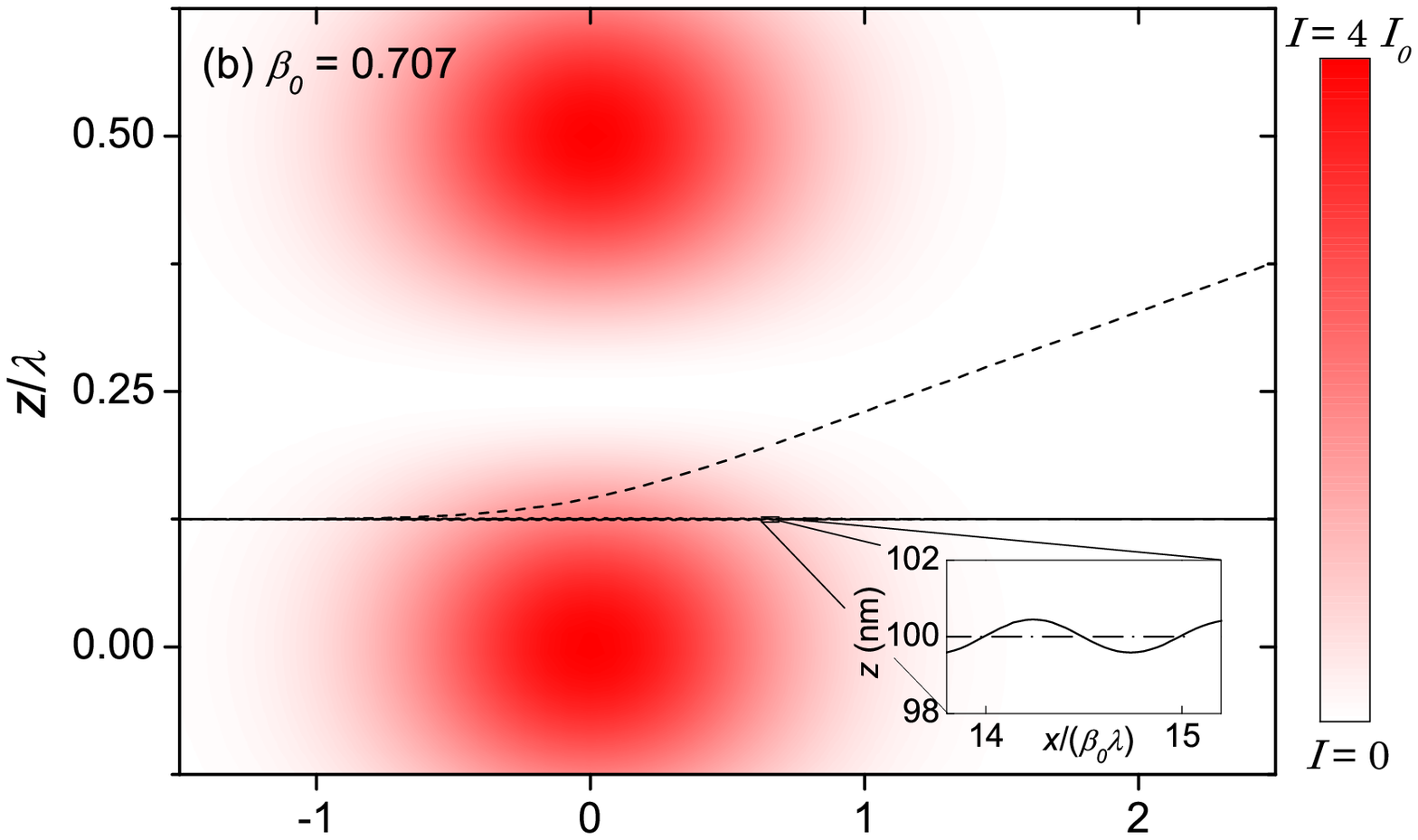}\\[-2mm]
\includegraphics[trim = 0.1in 0in 0.6in 0in, clip, width=0.9\columnwidth]{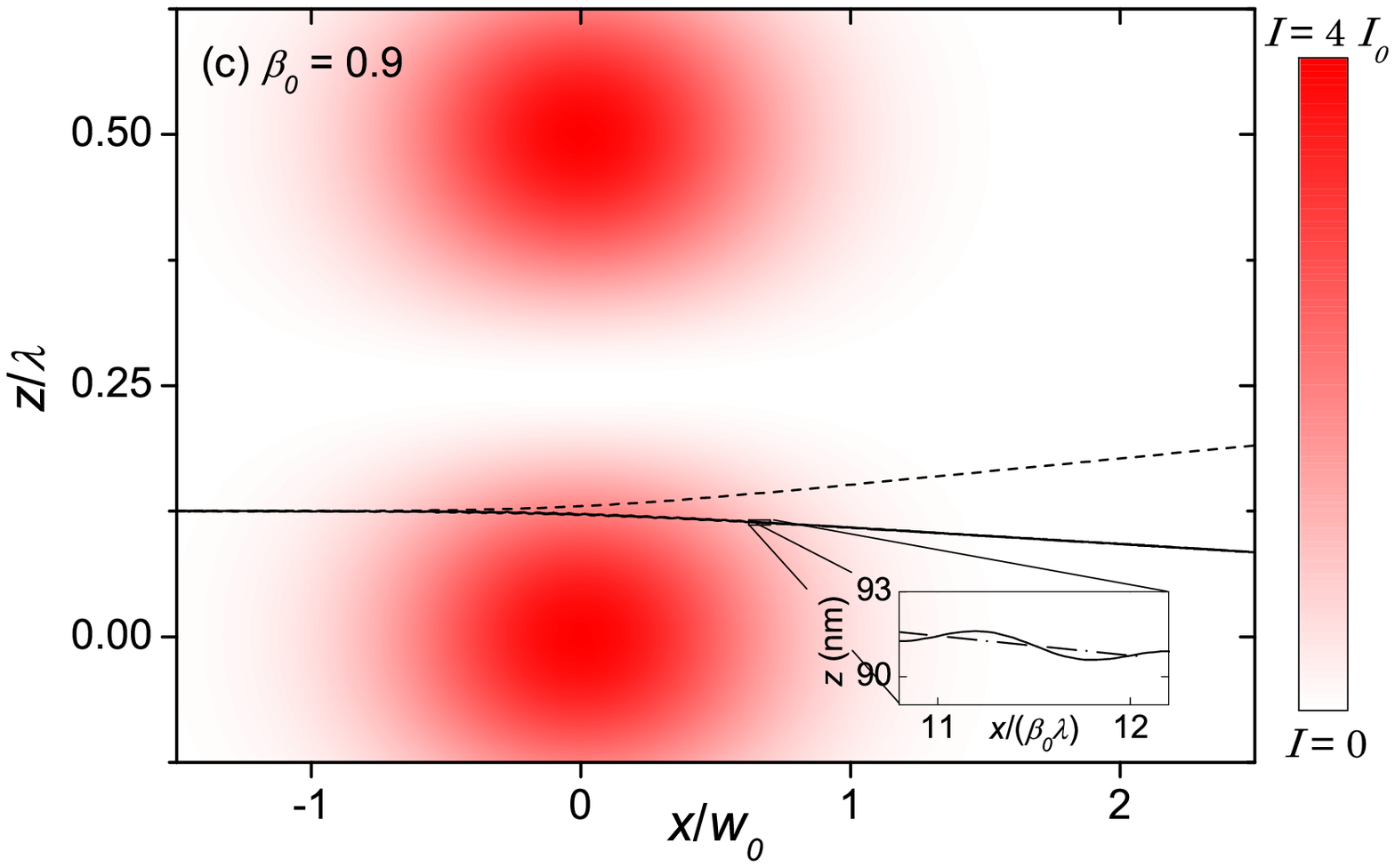}
\end{tabular}
\caption{\label{fig3} Trajectory of an electron incident on a standing wave that is polarized in the $x$-direction (left-right), for an initial velocity $\beta_0$ equal to (a) $0.5$; (b) $0.707\approx 1/\sqrt{2}$; (c) $0.9$. The color map shows the field intensity of the standing wave in the $(z,x)$-plane at time $t=0$. In each plot, the trajectory has been calculated with the methods GPT (solid line), OLD (dashed line) and NEW (dash-dotted line).}
\end{figure}

In Figs. \ref{fig2} and \ref{fig3}, the trajectory of the electron is shown as viewed from the negative $y$-axis, for several values of $\beta_0$ and for two different polarization directions. The trajectories are calculated using three different methods:
\begin{itemize*}
\item{(GPT) numerical integration of the exact equations of motion using the GPT code (solid lines);}
\item{(OLD) calculation according to the classical ponderomotive force, Eq. (\ref{1}) (dashed lines);}
\item{(NEW) calculation according to our polarization-dependent ponderomotive force, Eq. (\ref{22}), with $\overline{\bm{A}}_\perp$ replaced by $\overline{\bm{E}}_\perp/\omega$ (dash-dotted lines).}
\end{itemize*}
Standard paraxial field expressions have been used \cite{Quesnel}; these are listed in Appendix \ref{app.3} for reference.

Fig. \ref{fig2} shows the configuration in which the polarization is perpendicular to the initial velocity. In this case the polarization-dependent terms of Eq. (\ref{22}) vanish, and Eq. (\ref{22}) reduces to the classical expression, Eq. (\ref{1}). Indeed, for all three initial velocities the three descriptions yield identical trajectories to within the width of the lines, showing that the classical expression gives an excellent description of the averaged motion of the electron. The usual behavior can be seen in which the electron is deflected towards low-intensity regions.

In Fig. \ref{fig3}, however, the situation is very different. Here, the polarization is parallel to the initial velocity, so that the polarization-dependent terms of Eq. (\ref{22}) become important. Because of this, in Fig. \ref{fig3}(a) the magnitude of the ponderomotive force is smaller than in the corresponding case of Fig. \ref{fig2} ($\beta_0=0.5$), the ponderomotive force vanishes in Fig. \ref{fig3}(b), and it even changes direction towards the high-intensity region in Fig. \ref{fig3}(c). In all of these cases, the resulting trajectories are excellently predicted by Eq. (\ref{22}). Fig. \ref{fig3} demonstrates the relativistic reversal described in Ref. \cite{Pokrovsky} and Section \ref{sec.3.1}. Meanwhile, Eq. (\ref{1}) is insensitive to the polarization direction, so that the trajectories (dashed lines) are incorrectly predicted to be identical to the corresponding cases in Fig. \ref{fig2}.

The insets of Fig. \ref{fig3} are close-ups of the solid lines, showing that these actually consist of the smooth, time-averaged trajectory predicted by Eq. (\ref{22}) (dash-dotted line), and the GPT trajectory. The latter contains also the quiver motion, which has components in the $x$-direction (Eq. (\ref{18})) and the $z$-direction (Eq. (\ref{21})) and is therefore visible in the plane of drawing of Fig. \ref{fig3}. The frequency of this quiver motion is $\omega$ while the electron moves forward with a velocity $\beta_x\approx\beta_{0x}$, so that the spatial period of the quiver motion is $1/(\beta_{0x}\lambda)$. This has been made visible by scaling the $x$-axes of the insets. In Fig. \ref{fig2}, the quiver motion would not be visible in a close-up because it is perpendicular to the plane of drawing, since $\overline{\bm{A}}_\perp\cdot\bm{p}_{0\perp}=0$ in Eq. (\ref{21}) and the $z$-component of the quiver motion vanishes.\\

\begin{figure}[p]
\includegraphics[trim = 0in 0in 0in 0in, clip, width=0.8\columnwidth]{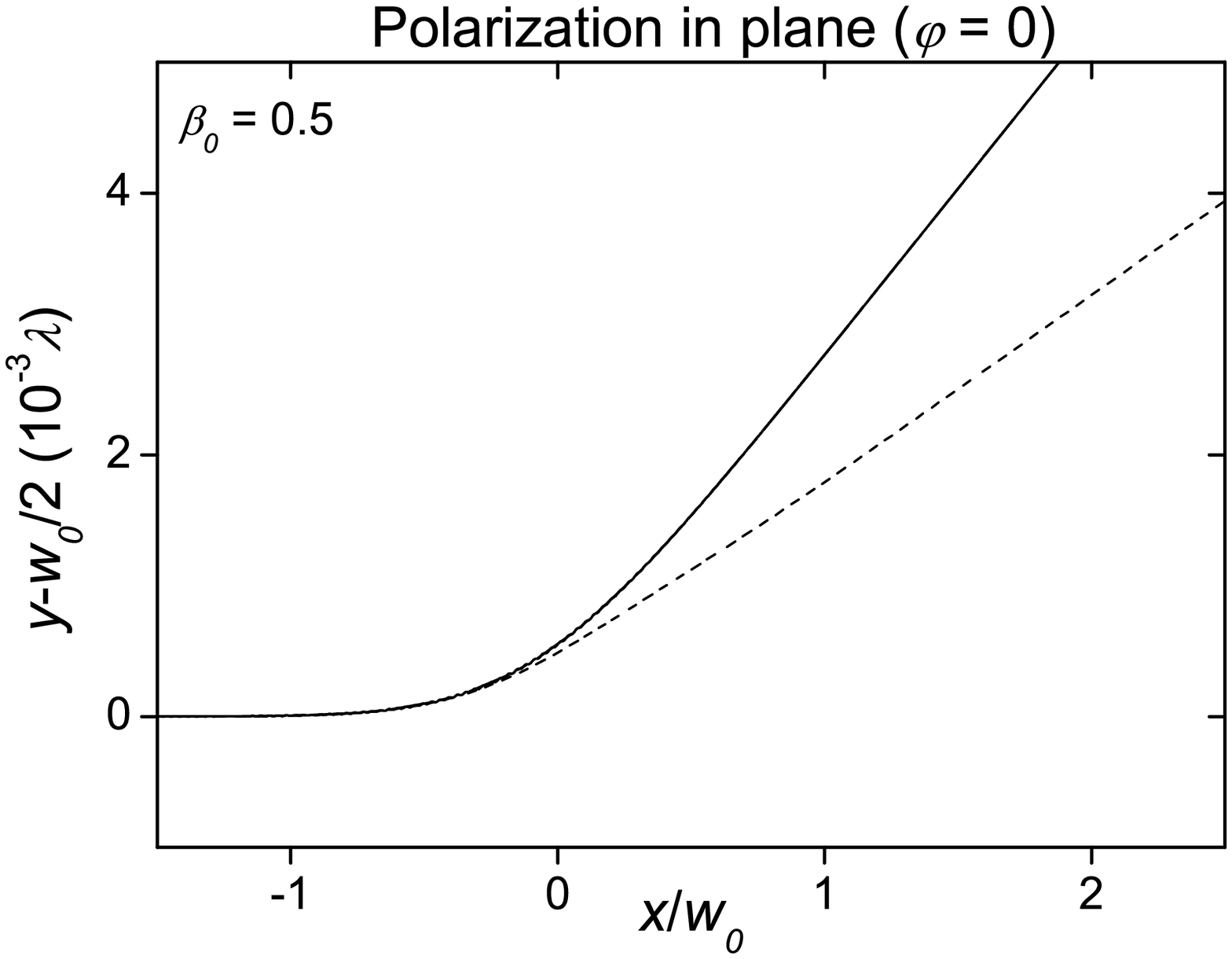}
\caption{\label{fig4}
Sideview from the positive $z$-axis of the trajectories shown in Fig. \ref{fig3}(a). The trajectories has been calculated with the methods GPT (solid line), OLD (dashed line) and NEW (overlapping with GPT).}
\end{figure}

In order to show some ponderomotive effects in the perpendicular direction as well, in Fig. \ref{fig4} a sideview is given of the trajectories of Fig. \ref{fig3}(a), as seen from the positive $z$-axis. Note from the vertical scale that in this direction the deflection of the electron is very small due to the very small field gradient. Nevertheless, again it is clear from the figure that Eq. (\ref{22}) accurately predicts the electron trajectory, contrary to the classical expression. Thus Eq. (\ref{22}) gives a precise description of the three-dimensional electron trajectory in a realistic, non-idealized field configuration. We have also repeated the simulations for combinations of $\epsilon$ and $a$ other than $a\sim\epsilon$. Whenever both $\epsilon$ and $a$ are less than about 0.1, consistent with the assumptions $a,\epsilon\ll 1$ made in Section \ref{sec.2}, we find the same level of agreement with the GPT results.

\section{Conclusion}
The classical polarization-independent ponderomotive force expression is commonly used to describe the time-averaged motion of a charged particle in an inhomogeneous oscillating EM field. It is generally assumed that this is an accurate description, at least for nonrelativistic field intensities. However, we have shown that this is not always true. If the field configuration possesses a direction in which the field changes on the scale of the wavelength, i.e. in a standing wave, and in addition the charged particle is relativistic, the ponderomotive force is modified. In particular, it becomes dependent on the polarization of the field. Because of this, the ponderomotive force may even vanish, or change its direction toward high field regions, as was found earlier by Kaplan and Pokrovsky \cite{Kaplan}. We have derived the modified ponderomotive force expression for these configurations, which is of gradient form like the classical expression. Comparison with simulations in the case of a realistic, non-idealized, three-dimensional field configuration confirmed the general validity of the analytical results.\\

The modifications of the ponderomotive force derived in this paper may have important implications for applications that involve the ponderomotive interaction of relativistic charged particles and standing EM waves. For example, in the electron bunch length measurement based on ponderomotive scattering of the electrons by a standing wave \cite{Hebeisen2}, the polarization of the wave is essential for an optimal design of the experimental setup. In the proposed X-ray free electron laser relying on the wiggling motion of electrons induced by the ponderomotive force in a standing wave \cite{Balcou}, the frequency of wiggling and hence that of the stimulated radiation directly depend on the polarization. Experimental tests involving the controlled scattering of electrons by a standing wave have confirmed the classical ponderomotive force expression \cite{Bucksbaum} and Kapitza-Dirac diffraction \cite{Freimund} using nonrelativistic electrons. It would be very interesting to extend these experiments to relativistic electrons to test the polarization-dependent ponderomotive force, Eq. (\ref{22}), in the classical limit, and to study the polarization-dependence of Kapitza-Dirac diffraction.

\appendix
\section{Order equations\label{app.1}}
Substitution of the expansions (\ref{8})-(\ref{13}) in the equations of motion (\ref{6})-(\ref{7}), and collecting terms of equal order in $\epsilon$, results in the following order equations. The components of the vector potential and the spatial derivatives have been treated as $e\bm{A}_\perp/(mc)=Ord(\epsilon)$, $eA_z/(mc)=Ord(\epsilon^2)$, $\lambda\nabla_\perp=Ord(\epsilon)$ and $\lambda\partial/\partial z=Ord(1)$ consistent with Eq. (\ref{5}) and the assumption $a\sim\epsilon$. For reasons of clarity, the equations are displayed in dimensional form.
\\

\begin{widetext}

\begin{eqnarray}
Ord(1):\hspace{2.5cm}&&\nonumber\\
\frac{d\widetilde{\bm{p}}^{(0)}}{dt}&=&\bm{0};\label{A1}\\
\frac{d\left(\overline{\bm{x}}+\widetilde{\bm{x}}\right)^{(0)}}{dt}&=&\frac{\left(\overline{\bm{p}}+\widetilde{\bm{p}}\right)_\perp^{(0)}}{m\gamma^{(0)}}.\label{A2}\\
Ord(\epsilon):\hspace{2.5cm}&&\nonumber\\
\frac{d\widetilde{\bm{p}}_\perp^{(1)}}{dt}+e\frac{d\bm{A}_\perp}{dt}&=&\bm{0};\label{A3}\\
\frac{d\left(\overline{p}_z^{(0)}+\widetilde{p}_z^{(1)}\right)}{dt}&=&\frac{e}{m\gamma^{(0)}}\frac{\partial\overline{\bm{A}}_\perp}{\partial z}\cdot\left(\overline{\bm{p}}+\widetilde{\bm{p}}\right)_\perp^{(0)};\label{A4}\\
\frac{d\left(\overline{\bm{x}}+\widetilde{\bm{x}}\right)^{(1)}}{dt}&=&\frac{\left(\overline{\bm{p}}+\widetilde{\bm{p}}\right)^{(1)}}{m\gamma^{(0)}}-\frac{\left(\overline{\bm{p}}+\widetilde{\bm{p}}\right)^{(0)}\cdot\left(\overline{\bm{p}}+\widetilde{\bm{p}}\right)^{(1)}}{(mc\gamma^{(0)})^2}\frac{\left(\overline{\bm{p}}+\widetilde{\bm{p}}\right)^{(0)}}{m\gamma^{(0)}}.\label{A5}\\
Ord(\epsilon^2):\hspace{2.5cm}&&\nonumber\\
\frac{d\left(\overline{\bm{p}}_\perp^{(0)}+\widetilde{\bm{p}}_\perp^{(2)}\right)}{dt}&=&\frac{e}{m\gamma^{(0)}}\left(\nabla_\perp\overline{\bm{A}}_\perp\right)\cdot\left(\overline{\bm{p}}+\widetilde{\bm{p}}\right)_\perp^{(0)};\label{A6}\\
\frac{d\left(\overline{p}_z^{(1)}+\widetilde{p}_z^{(2)}\right)}{dt}+e\frac{dA_z}{dt}&=&\frac{e}{m\gamma^{(0)}}\left[\widetilde{z}^{(1)}\frac{\partial^2\overline{\bm{A}}_\perp}{\partial z^2}\cdot\left(\overline{\bm{p}}+\widetilde{\bm{p}}\right)_\perp^{(0)}+\frac{\partial\overline{\bm{A}}_\perp}{\partial z}\cdot\left(\overline{\bm{p}}+\widetilde{\bm{p}}\right)_\perp^{(1)}+\frac{\partial\overline{A}_z}{\partial z}\left(\overline{p}+\widetilde{p}\right)_z^{(0)}-\right.\nonumber\\
&&\left.\qquad\qquad-\frac{\left(\overline{\bm{p}}+\widetilde{\bm{p}}\right)^{(0)}\cdot\left(\overline{\bm{p}}+\widetilde{\bm{p}}\right)^{(1)}}{(mc\gamma^{(0)})^2}\frac{\partial\overline{\bm{A}}_\perp}{\partial z}\cdot\left(\overline{\bm{p}}+\widetilde{\bm{p}}\right)_\perp^{(0)}\right].\label{A7}\\
Ord(\epsilon^3):\hspace{2.5cm}&&\nonumber\\
\frac{d\left(\overline{\bm{p}}_\perp^{(1)}+\widetilde{\bm{p}}_\perp^{(3)}\right)}{dt}&=&\frac{e}{m\gamma^{(0)}}\left[\widetilde{z}^{(1)}\left(\nabla_\perp\frac{\partial\overline{\bm{A}}_\perp}{\partial z}\right)\cdot\left(\overline{\bm{p}}+\widetilde{\bm{p}}\right)_\perp^{(0)}+\left(\nabla_\perp\overline{\bm{A}}_\perp\right)\cdot\left(\overline{\bm{p}}+\widetilde{\bm{p}}\right)_\perp^{(1)}+\left(\nabla_\perp \overline{A}_z\right)\left(\overline{p}+\widetilde{p}\right)_z^{(0)}-\right.\nonumber\\
&&\left.\qquad\qquad-\frac{\left(\overline{\bm{p}}+\widetilde{\bm{p}}\right)^{(0)}\cdot\left(\overline{\bm{p}}+\widetilde{\bm{p}}\right)^{(1)}}{(mc\gamma^{(0)})^2}\left(\nabla_\perp\overline{\bm{A}}_\perp\right)\cdot\left(\overline{\bm{p}}+\widetilde{\bm{p}}\right)_\perp^{(0)}\right].\label{A8}
\end{eqnarray}

\end{widetext}

\section{General field strength \label{app.2}}
The ponderomotive force, Eq. (\ref{22}), has been derived under the assumption that $a\sim\epsilon$. Since this expression is the result of balancing terms of equal order in the equations of motion, one might expect that it would be affected by changing the order of magnitude of the vector potential to, for instance, $a\sim\epsilon^2$. This is not the case, however. We only give a sketch of the generalized derivation for arbitrary $a\ll 1$.\\

Repeating first the order expansion method for the case $a\sim\epsilon^n$, $n\geq 1$, it is not difficult to find the lowest order slowly varying term and the lowest order rapidly varying term of the expansions of $\bm{p}$ and $1/\gamma$. Also, the first two terms of the expansion of $\bm{A}$ follow straightforwardly. The right-hand side of the equation of motion (\ref{6}) is then formed by factoring out the product of these three expansions. Taking the time average of the result, it is found that the lowest order terms that are nonzero on average are precisely those that form the ponderomotive force given by Eq. (\ref{22}). In terms of the corresponding set of order equations analogues to those in Appendix \ref{app.1}, the first $2n-1$ orders of the momentum equations yield zero right-hand sides upon averaging, while the $2n$-th and $(2n+1)$-th orders evaluate to respectively the $z$-component and perpendicular component of Eq. (\ref{22}).\\

Also the opposite situation in which $1\gg a\sim\epsilon^{1/n}$, $n\geq1$, is possible. Since in this case factors of $\bm{A}$ in the equations of motion lead to terms of fractional order in $\epsilon$, it is appropriate to expand all quantities in power series in terms of $\epsilon^{1/n}$ rather than $\epsilon$. This is effected by using the same power series expansions as before, with the understanding that $\bm{p}^{(i)}=Ord(\epsilon^{i/n})$ rather than $\bm{p}^{(i)}=Ord(\epsilon^i)$, for example. Except for this modification, the derivation of the ponderomotive force is analogous to that for the case $a\sim\epsilon^n$ considered above, and again Eq. (\ref{22}) is found. Thus Eq. (\ref{22}) is valid for arbitrary $a\ll1$.

\section{Field expressions used in numerical calculations \label{app.3}}
The solid lines in Figs. \ref{fig2}-\ref{fig4} have been calculated by numerical integration of the equations of motion
\begin{eqnarray*}
\frac{d\bm{p}}{dt}&=&e\left(\bm{E}+\frac{1}{\gamma}\bm{p}\times\bm{B}\right);\\
\frac{d\bm{x}}{dt}&=&\frac{\bm{p}}{m\gamma},
\end{eqnarray*}
using for $\bm{E}$ and $\bm{B}$ the following paraxial Gaussian beam fields \cite{Quesnel}. For polarization in the $x$-direction ($\phi=0$),
\begin{eqnarray}
\bm{E}&=&\bm{E}_+\exp\left(-\frac{(z+ct)^2}{4(c\sigma)^2}\right)+\bm{E}_-\exp\left(-\frac{(z-ct)^2}{4(c\sigma)^2}\right);\nonumber\\
&&\label{B1}\\
\bm{B}&=&\bm{B}_+\exp\left(-\frac{(z+ct)^2}{4(c\sigma)^2}\right)+\bm{B}_-\exp\left(-\frac{(z-ct)^2}{4(c\sigma)^2}\right);\nonumber\\
\bm{E}_\pm&=&E_0\frac{w_0}{w}\left(\bm{e}_x\cos\psi_\pm\pm\frac{xw_0}{z_Rw}\bm{e}_z\sin\chi_\pm\right);\nonumber\\
\bm{B}_\pm&=&\frac{E_0}{c}\frac{w_0}{w}\left(\mp\bm{e}_y\cos\psi_\pm-\frac{yw_0}{z_Rw}\bm{e}_z\sin\chi_\pm\right),\nonumber
\end{eqnarray}
in which $E_0=\sqrt{2I_0/(\epsilon_0c)}$ is the peak electric field amplitude, $w=w_0\sqrt{1+z^2/z_R^2}$ is the beam waist, $z_R=kw_0^2/2$ is the Rayleigh length, and the Gouy phases are
\begin{eqnarray*}
\psi_\pm&=&\omega t\pm\left(kz-\arctan\frac{z}{z_R}+\frac{z}{z_R}\frac{x^2+y^2}{w^2}\right);\\
\chi_\pm&=&\omega t\pm\left(kz-2\arctan\frac{z}{z_R}+\frac{z}{z_R}\frac{x^2+y^2}{w^2}\right).\\
\end{eqnarray*}
The dashed lines in Figs. \ref{fig2}-\ref{fig4} have been calculated according to Eq. (\ref{1}) with $I(\bm{x},t)=\epsilon_0c\langle\overline{E}_x^2\rangle$ using Eq. (\ref{B1}) for $\overline{E}_x$. The dash-dotted lines have been calculated according to Eq. (\ref{22}) with $\overline{\bm{A}}_\perp$ replaced by $\overline{E}_x/\omega$. For polarization in the $y$-direction ($\phi=\pi/2$), replace $x\rightarrow y$ and $y\rightarrow-x$ in the expressions above.

\end{document}